\documentclass[aps,prl,twocolumn,superscriptaddress,nobibnotes,10pt]{revtex4-2}
\usepackage{bm}
\usepackage{amsmath,amssymb}
\usepackage{graphicx}

\DeclareMathOperator{\sgn}{sgn}

\DeclareMathOperator{\Sp}{\mathrm{Sp}}

\begin{document}

\title{Landau-Zener tunneling and quantum interference of Andreev states}
\author{Mikhail S. Kalenkov}
\affiliation{I.E.Tamm Department of Theoretical Physics, P.N.Lebedev Physical Institute, 119991 Moscow, Russia}
\author{Andrei D. Zaikin}
\affiliation{I.E.Tamm Department of Theoretical Physics, P.N.Lebedev Physical Institute, 119991 Moscow, Russia}
\affiliation{National Research University Higher School of Economics, 101000 Moscow, Russia}

\date{\today}
\begin{abstract}
With the aid of a microscopic theory we derive an effective Hamiltonian that controls quantum dynamics of Andreev states in superconducting nanojunctions out of equilibrium. Resolving the corresponding Schr\"odinger-like equation we obtain the "wave functions" for Andreev levels and evaluate electric current across the junction in the presence of an external voltage bias. We demonstrate that quantum interference of Andreev states may yield pronounced coherent oscillations of the supercurrent as a function of the Josephson phase in junctions with barrier transmissions slightly below unity. Further implications of this novel effect are expected for junctions with diffusive barriers.
\end{abstract}
\maketitle

{\it Introduction.} Josephson effects \cite{Jos} in superconducting junctions with low transmission tunnel barriers are well understood \cite{BP}.
The same applies to equilibrium properties of superconducting weak links at arbitrary transmissions \cite{KO,KGI}. The situation becomes more complicated if the system is driven out of equilibrium, e.g., by an externally applied bias voltage $V$. Theoretical analysis of ac Josephson effect is successful in junctions at full transmissions \cite{Uwe,AB1,AB2} demonstrating qualitatively new features as compared to the standard tunneling limit \cite{BP}.

What if the barrier transmission is only slightly below unity? An attempt to answer this question was made by Averin and Bardas \cite{AB1} employing a physical picture of Landau-Zener tunneling between two subgap Andreev levels schematically illustrated in Fig. 1. In this way they evaluated the time-dependent supercurrent $I$ across the junction in the form \cite{AB1}

\begin{figure}
\begin{center}
\includegraphics[width=8cm]{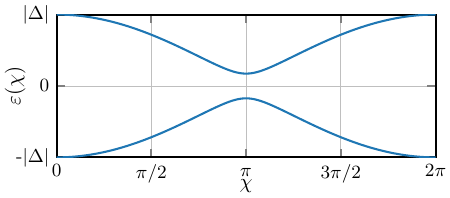}
\end{center}
\caption{Phase dependent energy of subgap Andreev levels in superconducting junctions at high transmissions.}
\label{ABS-fig}
\end{figure} 

\begin{equation}
I
= I_c\sin \Bigl[\frac{\chi }{2}\Bigr] \sgn V
\begin{cases}
1, & 0<\chi <\pi,
\\
[2e^{-\frac{\pi R|\Delta |}{eV}}-1], & \pi<\chi <2\pi,
\end{cases}
\label{Av}
\end{equation}
where $I_c$ is the junction critical current, $R \ll 1$ is its barrier reflection coefficient, $\Delta$ denotes the superconducting order parameter and $\chi (t)=2eVt$ defines the time dependent Josephson phase difference in the presence of a constant voltage $V$ applied to the junction. The current-phase relation (CPR) \eqref{Av} naturally reduces to that at full transmissions \cite{AB1,AB2} in the limit $R \to 0$ and to the low temperature equilibrium result \cite{KO} in the limit $V \to 0$. Here and below we set Planck constant $\hbar$ equal to unity.

More recently, Galaktionov and one of the authors \cite{GZ1} reconsidered the problem by performing a regular perturbative expansion in $R \ll 1$ within the standard physical picture involving the concept of multiple Andreev reflections (MAR) \cite{MAR}. While the time average current value $\overline{I}$ following from the formula \eqref{Av} was reproduced, CPR obtained within the applicability range of the perturbation theory  \cite{GZ1} turned out different from that in Eq. \eqref{Av}.

Turning back to the issue of Landau-Zener tunneling, we point out that its standard quantum mechanical treatment is not sufficient here at least for two reasons. Firstly, Andreev states correspond to a non-trivial superposition of quasiparticles and holes and, hence, cannot in general be treated as ordinary single particle levels. Secondly, in the presence of any non-zero voltage bias Andreev states demonstrate {\it non-unitary} time evolution \cite{KZ} generally implying both dissipation and decoherence.

In this Letter we are going to microscopically derive an effective Hamiltonian and formulate a Schr\"odinger-like equation that allows to successfully overcome the abovementioned problems. Within the framework of this physical picture -- which is complementary to (and fully consistent with) that of MAR -- we will evaluate electric current across the system and demonstrate that at low but non-zero values of $R$ CPR of a voltage-biased superconducting junction may exhibit pronounced coherent oscillations caused by quantum interference between Andreev states.

{\it The model and effective Hamiltonian.} 
We will consider the standard model for a single channel superconducting junction with transmission $D=1-R$ biased by an external voltage $V(t)$. This problem can be treated microscopically within the framework of the quasiclassical Eilenberger-Keldysh equations \cite{BWBSZ} supplemented by Zaitsev boundary conditions \cite{Zaitsev}. As a result, one arrives at a formally exact expression for electric current, which can be written in a compact form \cite{KZ}
\begin{multline}
I(t)=
ie
\Sp\Bigl\{
[\tilde W +  a ]^{-1}
[h - a h a^+]
\left([\tilde W +  a ]^{-1}\right)^+
\\
\times\tilde W^+[\chi(t)] \partial_{\chi} \tilde W[\chi(t)]
\Bigr\},
\label{T3}
\end{multline}
where all products are understood as time convolutions,
$a(\varepsilon) = -\varepsilon/|\Delta| + \sqrt{(\varepsilon/|\Delta|)^2 - 1}$ is the energy dependent Andreev amplitude, 
$h(\varepsilon) = \tanh(\varepsilon/2T)$, and we defined
\begin{equation}
\tilde W(\chi)  =
\begin{pmatrix}
W_d(\chi) & W_o(\chi) \\
-W_o(\chi) & W_d(\chi)
\end{pmatrix}
\label{tW}
\end{equation}
with $W_d =\sqrt{D} \cos (\chi/2) \hat p_3+ \sqrt{R}\hat p_1$, $W_o = \sqrt{D} \sin (\chi/2) \hat 1$. Here and below $\hat p_{1,2,3}$ are Pauli matrices and $\hat1$ is $2\times2$ unity matrix.

 The task at hand is to evaluate the inverse operator $(\tilde W +  a)^{-1}$ expressed in the following exact form
\begin{multline}
(\tilde W +  a)^{-1}
=
-\dfrac{|\Delta|}{2}
\begin{pmatrix}
(\varepsilon - \mathcal{H})^{-1} & 0 \\
0 & (\varepsilon - \mathcal{H})^{-1}
\end{pmatrix}
\\
\times\Biggl[
\begin{pmatrix}
W_o a^{-1} W_o^{-1} & 0
\\
0 & W_o a^{-1} W_o^{-1}
\end{pmatrix} \tilde W^+ + 1
\Biggr] 
\label{inv2}
\end{multline}
with an effective Hamiltonian
\begin{gather}
\mathcal{H} =
\dfrac{|\Delta|}{2}
\Bigl[
W_d
+
W_o a^{-1}  W_o^{-1} W_d a
+
W_o a^{-1}  W_o^{-1}
-
a^{-1}
\Bigr].
\label{Heff}
\end{gather}
It will be convenient for us to restrict our analysis to the Josephson phase values within the interval $0 <\chi (t) < 2\pi$. With sufficient accuracy (see End Matter) the expression for the Hamiltonian \eqref{Heff} can be reduced to
\begin{gather}
\mathcal{H} =
\mathcal{H}_0
+
\dfrac{i\dot \chi(t)}{4}\hat 1
\cot \Bigl[\frac{\chi(t)}{2}\Bigr]
-
\dfrac{\dot \chi(t)}{4}
\hat p_3
\label{Heff3}
\end{gather}
with $\mathcal{H}_0 =|\Delta|W_d$ and $\dot \chi(t)=2eV(t) \ll |\Delta |$.

In order to evaluate the operator $(\varepsilon - \mathcal{H})^{-1}$ let us obtain a general solution of the equation $i \partial \Psi/\partial t = \mathcal{H} \Psi$. Making use of a substitution 
\begin{equation}
\Psi =\sqrt{\sin [\chi(t)/2]}\tilde \Psi,
\label{subst}
\end{equation}
we arrive at a Schr\"odinger-like equation
\begin{equation}
i \dfrac{\partial \tilde \Psi}{\partial t}
=\tilde{\mathcal{H}}
\tilde \Psi, \quad
\tilde{\mathcal{H}}
=
\mathcal{H}_0
-
\dfrac{\dot \chi(t)}{4}\hat p_3
\label{shrod}
\end{equation}
with the Hermitian effective Hamiltonian $\tilde{\mathcal{H}}$. Note that linear in $\dot \chi(t)$ terms must be retained here: Although small, they accumulate over long times leading to non-negligible changes of the "wave function". We also point out that -- although technically our formalism does not require this limitation -- for the sake of simplicity below we restrict our analysis to the limit of sufficiently slowly varying in time bias voltage which -- if one so wishes -- can then be replaced by a constant $V(t)=V$.

It is convenient to introduce two eigenvectors of the matrix $\mathcal{H}_0$ corresponding to two subgap Andreev bound 
states with energies
$\pm \varepsilon_A(\chi)$
\begin{equation}
\mathcal{H}_0(\chi) \varphi_{\pm}(\chi) = \pm \varepsilon_A(\chi)\varphi_{\pm}(\chi), 
\label{H0}
\end{equation}
where $\varepsilon_A(\chi)=|\Delta|\sqrt{1 - D \sin^2(\chi/2)}$.
The eigenvectors $\varphi_{\pm}(\chi)$ can be identified in terms of a single real vector
\begin{equation}
\varphi(\chi)
=
\dfrac{
\begin{pmatrix}
\sqrt{R} |\Delta | \\
\varepsilon_A(\chi) - \sqrt{D} |\Delta | \cos (\chi/2)
\end{pmatrix}}{
\sqrt{2 \varepsilon_A(\chi) \left[ \varepsilon_A(\chi) - \sqrt{D} |\Delta | \cos (\chi/2)\right]}
}
\label{vector}
\end{equation}
by means of the equations
$\varphi_{+} = \varphi(\chi)$ and $\varphi_{-} = i \hat  p_2 \varphi(\chi)$. Far from the anticrossing point  $\chi(t_{\pi}) = \pi$ Eq. \eqref{shrod} can  be solved within an adiabatic approximation with the result
\begin{gather}
\psi_{\pm}(t) =
\dfrac{\varphi_{\pm}[\chi(t)]}{\sqrt{a[\pm\varepsilon_A[\chi(t)]]}} \exp\left(
\mp i \int_{t_{\pi}}^t\varepsilon_A[\chi(t')] d t'
\right)
\label{psit}
\end{gather}
Making use of Eq. \eqref{psit} it is possible to construct two orthogonal solutions of Eq. \eqref{shrod} describing the whole phase interval  including the anticrossing region $\chi \approx \pi$. These solutions read
\begin{gather}
\tilde\Psi_{\mu} (t) = \sum_{\nu=\pm}\mathcal{S}_{\nu,\mu}(t) \psi_{\nu}(t),
\quad
\tilde \Psi_{\mu}^+(t) \tilde \Psi_{\tilde \mu}(t) = \delta_{\mu, \tilde \mu}.
\label{rddef}
\end{gather}
Here $\mathcal{S}_{\nu,\mu}(t)$ are elements of the unitary scattering matrix
\begin{equation}
\mathcal{S}(t)=
\begin{pmatrix}
r(t) & d(t) \\
-d^*(t) & r^*(t)
\end{pmatrix}, \quad  |r(t)|^2 + |d(t)|^2 = 1
\label{S}
\end{equation}
with the functions $r(t)$ and $d(t)$ to be determined below. For later convenience, we also define the phase
\begin{equation}
\delta(t)=\pi-\arg d(t)-\arg r(t).
\label{deltat}
\end{equation}

Employing two linearly independent solutions $\tilde\Psi_{\mu} (t)$ of Eq. \eqref{shrod}, one readily finds the kernel of the operator
\begin{equation}
(\varepsilon - \mathcal{H})^{-1}
=
-i \theta(t - t')
\sum_{\mu =\pm}
\tilde \Psi_{\mu}(t)
\sqrt{\dfrac{\sin [\chi(t)/2]}{\sin [\chi(t')/2]}}
\tilde \Psi^+_{\mu}(t').
\label{invH}
\end{equation}
The integral kernel of $(\tilde W +  a)^{-1}$ then reads
\begin{multline}
(\tilde W +  a)^{-1}
=
i\dfrac{\sqrt{D}|\Delta|}{2}
\theta(t - t')
\sqrt{\sin [\chi(t)/2]\sin [\chi(t')/2]}
\\\times
\sum_{\mu, \nu, \tilde \nu}
\mathcal{S}_{\nu,\mu}(t) \psi_{\nu}(t)
\begin{pmatrix}
i  & -1 \\
1 & i
\end{pmatrix}
\dfrac{\mathcal{S}^*_{\tilde \nu,\mu}(t') \psi^+_{\tilde \nu}(t')}{a(\tilde \nu \varepsilon_A[\chi(t')])},
\label{inv11}
\end{multline}
where $\psi_{\nu}(t)  M  \psi^+_{\tilde \nu}(t')$ denotes a block matrix with entries $m_{ij} \psi_{\nu}(t)\psi^+_{\tilde \nu}(t')$.

Substituting Eq. \eqref{inv11} into Eq. \eqref{T3}, making use of the exact relation
$\Psi^*_{\mu}(t) = \mu\hat  p_2 \Psi_{-\mu}(t) $ (enabling one to perform a summation over $\mu$), fixing $d(t) \to 0$ for $\chi \ll \pi$ by an appropriate choice of $\tilde\Psi_{\pm}(t)$ and restricting ourselves to the low temperature limit $T\ll |\Delta |$ we arrive at the general expression for CPR out of equilibrium
\begin{equation}
I(t)=
\tilde \Psi^+_+ (t)\hat J[\chi (t)] \tilde \Psi_+ (t),
\label{T7}
\end{equation}
where we defined the matrix current operator
\begin{equation}
\hat J (\chi)=
e D |\Delta|
\sin (\chi/2)
\begin{pmatrix}
\sqrt{D} & \sqrt{R} e^{i\chi/2 }  \\
\sqrt{R}e^{-i\chi/2 } & -\sqrt{D}
\end{pmatrix}.
\label{A}
\end{equation}

{\it Andreev states evolution and Landau-Zener tunneling.} In order to evaluate the current $I(t)$ \eqref{T7} as a function of the phase $\chi (t)$ we now need to solve the Schr\"odinger-like equation \eqref{shrod} and to construct the ``wave function'' $\tilde\Psi_+ (t)$.  At low voltages $eV \ll |\Delta |$ and for the phase values $\chi (t)$ smaller than and not very close to $\pi$ one can set $d(t)=0$, which yields $\tilde\Psi_+(t) = \psi_+(t)$ in this region.

Provided the phase values $\chi (t)$ become sufficiently close to $\pi$ the adiabatic approximation breaks down due to strong coupling between Andreev levels. At low voltages, the corresponding nonadiabatic dynamics of the Andreev levels is realized for weakly reflecting junctions $R\ll1$. Introducing dimensionless time variable $\tau = \sqrt{|\Delta|\dot\chi(t_{\pi})} (t-t_{\pi}) + (1/2)\sqrt{\dot\chi(t_{\pi})/|\Delta|}$
and the adiabadicity parameter $s = R|\Delta|/\dot\chi(t_{\pi})$, close to the point $\chi =\pi$
we can reduce Eq. \eqref{shrod} to the form 
\begin{equation}
i \dfrac{\partial \tilde \Psi}{\partial \tau}
=
\begin{pmatrix}
-\tau/2  & \sqrt{s} \\
\sqrt{s} & \tau/2
\end{pmatrix}
\tilde \Psi ,
\label{shrod3}
\end{equation}
which is standard for the problem of Landau-Zener tunneling. The solution of this problem is well known \cite{Z32} and allows one to immediately determine the scattering matrix $\mathcal{S}$ \eqref{S}. In particular, for $\chi > \pi$ and sufficiently far from the anticrossing point $\chi=\pi$ one finds \cite{Z32,Vit96,Iv23}
\begin{gather}
\label{d}
d = - e^{-\pi s},
\\
\delta =
 -\frac{\pi}{4} + s\ln \frac{s}{e}  -  \arg \Gamma(is),
\label{delta}
\end{gather}
where $\Gamma (x)$ is the Euler gamma-function.

{\it CPR and interference of Andreev states.} Now we are ready to evaluate the supercurrent $I$ across the junction. Combining Eqs.
\eqref{psit}-\eqref{deltat} with Eqs. \eqref{T7}-\eqref{A}, we get
\begin{multline}
I(t)=-2e
\dfrac{\partial \varepsilon_A(\chi)}{\partial \chi}\Biggl[
1 - 2|d(t)|^2
+
2  \sqrt{R} |r(t)| |d(t)|
\\\times
\tan [\chi(t)/2]
\cos\left( \delta(t)  + 2 \int_{t_{\pi}}^t\varepsilon_A[\chi(t')] d t' \right)
\Biggr].
\label{T+}
\end{multline}
Equation \eqref{T+} together with Eqs. \eqref{d}, \eqref{delta} represents the central result of our present work. It establishes a general relation between the current $I(t)$ and the time dependent Josephson phase $\chi (t)$ in superconducting junctions at sufficiently low bias voltages $eV \ll |\Delta |$ and low temperatures.

Let us for a moment restrict ourselves to a simple step-like approximation for the probability of Landau-Zener tunneling 
\begin{equation}
|d|^2\equiv 1-|r|^2=
\begin{cases}
0, & 0<\chi <\pi,
\\
e^{-\frac{\pi R|\Delta |}{eV}}, & \pi <\chi <2\pi
\end{cases}
\label{d2}
\end{equation}
and compare our result \eqref{T+} with that in Eq. \eqref{Av} \cite{AB1}. The key difference between these results lies in the presence of the oscillating contribution to the supercurrent $\propto |r||d|$ (second and third lines in Eq. \eqref{T+}) at  $\pi <\chi <2\pi$ which is totally absent in Eq. \eqref{Av}. The appearance of this contribution is easy to understand: After Landau-Zener tunneling event both Andreev states get partially occupied and exhibit quantum interference which manifests itself in the above oscillations. As follows from Eq. \eqref{T+}, the amplitude of these oscillations reaches its maximum at $|r|\approx |d| \approx 1/\sqrt{2}$ and vanishes in both limits $|r| \to 0$ and $|d| \to 0$. Thus, we predict a novel effect of coherent oscillations of the supercurrent due to quantum interference of Andreev states in superconducting junctions driven by a
(small) external voltage bias $V$. 

\begin{figure}
\begin{center}
\includegraphics[width=8cm]{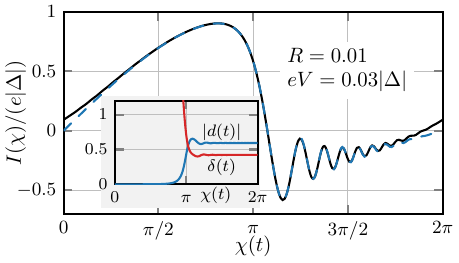}
\end{center}
\caption{The current $I$ in a single channel junction as a function of $\chi(t)$ for  $R=0.01$ and $eV=0.03 |\Delta|$. Black solid line represents a numerically exact result, while dashed blue line corresponds to Eq. \eqref{T+} with $r(t)$ and $d(t)$ obtained from the solution of the Zener problem (see End Matter). Inset: exact dependencies $d(t)$ and $\delta(t)$.}
\label{Ichi-R001-V003-fig}
\end{figure}

\begin{figure}
\begin{center}
\includegraphics[width=8cm]{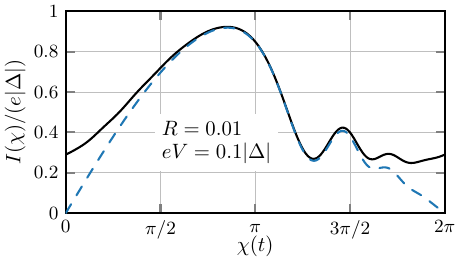}
\end{center}
\caption{The same as in the main panel of Fig.~\ref{Ichi-R001-V003-fig} for $eV = 0.1 |\Delta|$.}
\label{Ichi-R001-V010-fig}
\end{figure}

\begin{figure}
\begin{center}
\includegraphics[width=8cm]{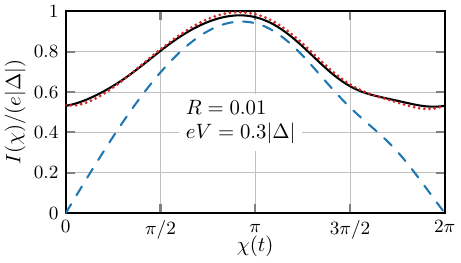}
\end{center}
\caption{The same as in the main panel of Fig.~\ref{Ichi-R001-V003-fig} for $eV = 0.3 |\Delta|$. The red dotted line represents the sum $I+\delta I$ defined by Eqs. \eqref{T+} and \eqref{diss}.}
\label{Ichi-R001-V030-fig}
\end{figure}

Note that the step-like approximation \eqref{d2} becomes insufficient in the vicinity of the anticrossing point $\chi =\pi$. In order to quantitatively describe the behavior $I[\chi (t)]$ close to this point it is necessary to perform a detailed calculation of the scattering matrix \eqref{S} which we relegate to End Matter. For the sake of comparison we also calculated CPR $I[\chi (t)]$ numerically employing exact recurrence relations \cite{KZ} equivalent to the expression \eqref{T3} in the limit of a constant in time bias voltage.
The numerical code used for these calculations is publicly available at \cite{Code}.
The corresponding numerically exact results are displayed in Figs. \ref{Ichi-R001-V003-fig}-\ref{Ichi-R001-V030-fig} together with the dependencies $I[\chi (t)]$ \eqref{T+} evaluated at $R=0.01$ and different values of $V$.

Figure \ref{Ichi-R001-V003-fig} demonstrates CPR at a small constant voltage value $V=0.03|\Delta |/e$ in which case the amplitude of coherent oscillations is close to its maximum at a given value of $R$ being only few times smaller than the critical current value $I_c=e|\Delta |$ for our junction. We indeed clearly observe these oscillations within the phase interval $\pi  < \chi (t) <2\pi$ with the amplitude
decreasing monotonously as one goes away from the anticrossing point $\chi =\pi$. The oscillation frequency $\omega$ follows adiabatically the instantaneous level splitting as $\omega[\chi(t)]=2\varepsilon_A[\chi(t)]$. We also observe practically ideal agreement between analytic and numerical results for CPR except for small deviations between the two curves close to the points $\chi =0$, $2\pi$.
CPR of a similar form is also recovered at even smaller values of $V$ (not shown), with oscillations of a relatively smaller amplitude and practically invisible difference between the curves in the vicinity of the phase values $\chi =0$, $2\pi$.

The same CPR dependencies evaluated both numerically and analytically at $V=0.1|\Delta |/e$ are displayed in Fig. \ref{Ichi-R001-V010-fig}. Coherent oscillations remain clearly visible at $\pi  < \chi (t) <2\pi$, although in a somewhat less pronounced form. A very good agreement between the two curves is again observed in the central part of the plot, though deviations between them towards the points $\chi =0,2\pi$  become somewhat bigger than in Fig. \ref{Ichi-R001-V003-fig}. This discrepancy increases further with increasing $V$, as it is demonstrated, e.g., in Fig. \ref{Ichi-R001-V030-fig}.

The reason for this discrepancy is as follows. So far we have completely neglected the sub-Ohmic dissipative contribution to the current $\delta I$ which at low voltages $eV \ll |\Delta |$ yields the terms $\propto V^{2/3}$ in the $I-V$ curve \cite{GZ2} and $\propto V^{1/3}$ in CPR \cite{KZ,GZ3}. Furthermore, this dissipative contribution is mostly concentrated in the vicinity of the phase values $\chi = 2\pi k$ \cite{FN}, just where the deviations between Eq. \eqref{T+} and the numerical curve are observed.

In order to account for this extra contribution to the current, at sufficiently small values of both $R$ and $|r|$ we may use the result \cite{GZ3} obtained for fully transmitting junctions. In our notations this result takes the form 
\begin{equation}
\delta I =\dfrac{be|\Delta|}{\pi}
\left(\dfrac{eV}{|\Delta|}\right)^{1/3}
\exp\left[
-0.27 b \left(\dfrac{|\Delta|}{eV}\right)^{1/3} |\delta\chi|
\right],
\label{diss}
\end{equation}
where $\delta\chi$ is the distance to the nearest point $2\pi n$, the parameter $b \simeq 2.5$  for $eV/|\Delta| \lesssim 0.01$ and $b \simeq 2.7 $  for $eV/|\Delta| \gtrsim 0.1$. In Fig. \ref{Ichi-R001-V030-fig} we also plotted the dependence $I(\chi)+\delta I(\chi)$ defined by Eqs. \eqref{T+} and \eqref{diss} which demonstrates a remarkable agreement with the numerical result at all values of $\chi$.

We also note that the parameter values in Fig. \ref{Ichi-R001-V030-fig} match with the applicability range of the perturbative calculation \cite{GZ1} $R|\Delta | \ll eV \ll 2|\Delta |$. The results derived here are fully consistent with those of Ref. \onlinecite{GZ1} within their applicability range. In particular, we point out that no coherent oscillations in CPR were recovered within the perturbative approach \cite{GZ1}. The reason for that becomes clear from our Eq. \eqref{T+}: Although the current oscillations are still present in the perturbative regime, their magnitude is controlled by the terms $\sim R\sqrt{|\Delta /(eV)}$. These terms are parametrically smaller than the leading correction to the current  $\sim R|\Delta |/(eV)$ and, hence, the current oscillations were essentially ignored in \cite{GZ1}. Accordingly, these oscillations are hardly visible also in our Fig. \ref{Ichi-R001-V030-fig}.

{\it CPR in diffusive junctions.} Our results can be directly extended to junctions with many conducting channels. Of particular interest are superconducting junctions with sufficiently short diffusive barriers in which case the reflection coefficients $R$ of conducting channels 
are controlled by the Dorokhov distribution \cite{Dor}
\begin{equation}
\rho(R) = \dfrac{\pi}{2 e^2 R_N} \dfrac{1}{(1-R)\sqrt{R}},
\label{Dor}
\end{equation}
where $R_N$ is the normal state resistance of a diffusive barrier. According to this bimodal distribution the barrier contains predominantly conducting channels with low and high transmission values. Averaging the expression for the current \eqref{T+} with the distribution \eqref{Dor} one should bear in mind that for the former group of channels it suffices to set $|d | \to 0$. Then one finds
\begin{multline}
I[\chi (t)]=\dfrac{\pi |\Delta |}{2eR_N}\Biggl\{\ln \left[\frac{1+\sin [\chi (t) /2]}{1-\sin [\chi (t)/2]}\right]\cos[\chi(t)/2]
\\+\sqrt{\dfrac{eV}{|\Delta|}}f[\chi (t)]+\dfrac{eV}{|\Delta|}g[\chi (t)]\Biggr\}.
\label{diff}
\end{multline}
The combination in the first line of Eq. \eqref{diff} represents the standard CPR for a diffusive superconducting junction in equilibrium at low temperatures $T \ll |\Delta |$ \cite{KO,ZP85} which -- as we demonstrated here -- also applies out of equilibrium at small bias voltages $eV \ll |\Delta |$. 

Additional contributions in the second line of Eq. \eqref{diff} originate from the terms in Eq. \eqref{T+} describing Landau-Zener tunneling
in which case only channels with $R \ll 1$ need to be taken into account. In particular, the contribution to the current $\propto \sqrt{V}$ in Eq. \eqref{diff} follows directly from the term $\propto |d|^2$ in Eq. \eqref{T+}. Accordingly, we have $f(\chi )=0$ for $0<\chi \lesssim \pi$ and $f(\chi )=2\sin (\chi /2)$ for $\pi\lesssim \chi <2\pi$ with a smooth crossover between these two regimes within a narrow region $|\chi-\pi|\sim \sqrt{|eV|/|\Delta|}$. The oscillating term in Eq. \eqref{T+} yields the last contribution $\propto V$ in Eq. \eqref{diff} and
$g(\chi )$ is an oscillating function containing terms proportional to $\cos \gamma (t)$ and $\sin\gamma (t)$ with $\gamma (t)=[2|\Delta|/(eV)]
[1 - \sin[(\chi(t)/2)]$.

Averaging Eq. \eqref{diff} over time, we obtain
\begin{equation}
\overline{I} =\dfrac{|\Delta|}{eR_N}\sqrt{\dfrac{eV}{|\Delta |}},
\end {equation}
i.e. at small voltages and $T \to 0$ the differential conductance of short diffusive superconducting junctions behaves as $d\overline{I}/dV \propto 1/\sqrt{V}$.

In summary, we microscopically derived an effective Schr\"odinger-like equation which describes quantum dynamics of Andreev states in superconducting nanojunctions out of equilibrium. By solving this equation one can find the "wave functions" corresponding to these states and evaluate electric current flowing across the junction. In the presence of an external voltage bias Landau-Zener tunneling occurs between Andreev levels. As a result, both Andreev states become occupied and start interfere with each other. Such quantum interference, in turn, may result in pronounced coherent oscillations of the current as a function of the time dependent Josephson phase $\chi (t)$. This effect has further implications for junctions with many conducting channels and can be directly tested in future experiments.

\section{End Matter}

\paragraph{Effective Hamiltonian.} Let us outline some technical details of our derivation of the effective Hamiltonian for quasiparticles occupying Andreev bound states in superconducting junctions at sufficiently low voltages. We start from the general expression for the current through a single conducting channel derived in Ref.~\cite{KZ}
\begin{multline}
I(t) = i e\Sp \Bigl\{
(\tilde{\mathbb{S}}_n  + a \hat \tau_1 )^{-1} (h - a h a^+)(\tilde{\mathbb{S}}_n^+  + a^+ \hat \tau_1)^{-1}
\\\times
\tilde{\mathbb{S}}_n^+[\chi(t)]
\partial_{\chi} \tilde{\mathbb{S}}_n[\chi(t)]
\Bigr\},
\label{It}
\end{multline}
where the functions $a$ and $h$ are defined in the main text. The matrices $\tilde{\mathbb{S}}_n[\chi(t)]$ and $\hat \tau_1$ are given by
\begin{equation}
\tilde{\mathbb{S}}_n(\chi)
=
\begin{pmatrix}
V(\chi) & 0 \\
0 & V^*(\chi) \\
\end{pmatrix},
\quad
\hat  \tau_1
=
\begin{pmatrix}
0 & 0 & 1 & 0 \\
0 & 0 & 0 & 1 \\
1 & 0 & 0 & 0 \\
0 & 1 & 0 & 0 \\
\end{pmatrix},
\label{tS}
\end{equation}
where
\begin{equation}
V(\chi) =
\begin{pmatrix}
\sqrt{R} & -i\sqrt{D} e^{-i\chi/2} \\
-i\sqrt{D} e^{i\chi/2} & \sqrt{R}
\end{pmatrix}.
\end{equation}

In Eq.~\eqref{It} and below all products are understood as convolutions in the time domain,
\begin{equation}
C=AB, \quad \Leftrightarrow \quad C(t,t') = \int A(t, \tilde t) B(\tilde t, t') d \tilde t.
\end{equation}
Provided an operator $A$ in the time representation is described by the kernel $A(t,t')$, the kernel of the operator $A^+$ reads $A^+(t', t)$. The inverse operator $B=A^{-1}$ has the kernel obeying the equation
\begin{equation}
B=A^{-1}, \quad \Leftrightarrow \quad \int d \tilde t B(t, \tilde t) A(\tilde t, t') = \delta(t - t').
\end{equation}
Below we will use the same notation for the operators both in time and energy representations related to each other as
\begin{equation}
A(t - t') = \int \dfrac{d \varepsilon}{ 2\pi} e^{-i\varepsilon(t-t')} A(\varepsilon).
\end{equation}
The choice of a particular representation will be clear from the context.

Performing a unitary transformation with a time-independent matrix in Eq.~\eqref{It}
\begin{equation}
\tilde  W(\chi)
=
U^+ \tilde{\mathbb{S}}_n(\chi) \hat \tau_1 U,
\quad
U =
\dfrac{1}{2}
\begin{pmatrix}
1 & i & i & 1 \\
1 & i & -i & -1 \\
i & 1 & 1 & i \\
i & 1 & -1 & -i
\end{pmatrix}
\label{UWU}
\end{equation}
we arrive at Eq.~\eqref{T3}.

Next we use the exact identity
\begin{multline}
\Biggl[
\begin{pmatrix}
W_o a^{-1} W_o^{-1} & 0
\\
0 & W_o a^{-1} W_o^{-1}
\end{pmatrix} \tilde W^+ + 1
\Biggr]
(\tilde W +  a)
\\=
-\dfrac{2}{|\Delta|}
\begin{pmatrix}
(\varepsilon - \mathcal{H}) & 0 \\
0 & (\varepsilon - \mathcal{H})
\end{pmatrix} ,
\label{inv22}
\end{multline}
where $\mathcal{H}$ is defined by Eq.~\eqref{Heff}. This identity can be verified directly by expanding both sides of Eq.~\eqref{inv22}, which immediately leads to the representation \eqref{inv2} for the inverse operator $(\tilde W+a)^{-1}$.

Here we stress that all the operators below are understood in the time domain as integral operators with kernels depending on the time differences. Formally, the function $F(\varepsilon)$ denotes an operator whose kernel in the time representation is given by the Fourier transformation
\begin{equation}
F(t-t') = \int \frac{d\varepsilon}{2\pi}\, e^{-i\varepsilon(t-t')} F(\varepsilon).
\end{equation}
In this sense, the variable $\varepsilon$ and any expression built from it should be understood as a shorthand notation for the corresponding convolution operators in the time domain. For compactness, we use the same notation $F(\varepsilon)$ both for the Fourier space functions and associated operators.

In order to proceed let us make use of the fact that in the low voltage limit the matrices $W_{d,o}$ are slowly varying functions of time. In this regime, we adopt a symbolic operator notation where the product $F(\varepsilon) f(t)$ implies a nonlocal operator acting on the function $f(t)$. With this convention, one can verify an approximate interchange rule
\begin{equation}
F(\varepsilon) f(t) \approx
f(t) F(\varepsilon) + i f'(t) F'(\varepsilon),
\label{inter}
\end{equation}
where $f(t)$ is a slowly varying function of time.

Making use of the identities for the Andreev reflection amplitude $a(\varepsilon)$,
\begin{gather}
a'/a = - \dfrac{1}{|\Delta|} \dfrac{2 }{a - 1/a} = -\dfrac{1}{\sqrt{\varepsilon^2 - |\Delta|^2}},
\\
a'/a^2 =  \dfrac{1}{|\Delta|}
\left(1 + \dfrac{\varepsilon}{\sqrt{\varepsilon^2 - |\Delta|^2}}\right)
\end{gather}
and employing the interchange rule \eqref{inter}, we arrive at the expression for the effective Hamiltonian
\begin{multline}
\mathcal{H} \approx
|\Delta|W_d
+
\dfrac{i}{2}
W'_o W_o^{-1}
\\+
\dfrac{i}{2}
\Bigl[
|\Delta|W'_d
+
W'_o W_o^{-1}  (\varepsilon - |\Delta| W_d)
\Bigr]
\dfrac{1}{\sqrt{\varepsilon^2 - |\Delta|^2}}.
\label{Heff2}
\end{multline}
Within the same approximation, one can replace $\varepsilon$ in the expression for $\mathcal{H}$ by $|\Delta|W_d$. This immediately yields the effective Hamiltonian \eqref{Heff3}.

\paragraph{Time-dependent Landau--Zener amplitudes.}
Here we present some technical details of our evaluation of the time-dependent Landau--Zener amplitudes $r(t)$ and $d(t)$ for the whole time domain. In the limit $R\ll1$ and $|\chi-\pi|\ll 1$, the functions $\psi_{\pm}(t)$ can be written (in dimensionless units) as
\begin{widetext}
\begin{gather}
\psi_+(t)
=
\dfrac{
\begin{pmatrix}
\sqrt{s} \\
\sqrt{s + (\tau - \tau_0)^2/4} + (\tau - \tau_0)/2
\end{pmatrix}}{
\sqrt{2 \sqrt{s + (\tau - \tau_0)^2/4} \left[\sqrt{s + (\tau - \tau_0)^2/4} + (\tau - \tau_0)/2\right]}
}
\exp\left(
-i\dfrac{\pi}{4}
-i  \Phi(\tau)
\right) ,
\label{psippi}
\\
\psi_-(t)
=
\dfrac{
\begin{pmatrix}
\sqrt{s + (\tau - \tau_0)^2/4} + (\tau - \tau_0)/2
\\
-\sqrt{s}
\end{pmatrix}}{
\sqrt{2 \sqrt{s + (\tau - \tau_0)^2/4} \left[\sqrt{s + (\tau - \tau_0)^2/4} + (\tau - \tau_0)/2\right]}
}
\exp\left(
-i\dfrac{\pi}{4}
+i \Phi(\tau)
\right) ,
\label{psimpi}
\end{gather}
where we define
\begin{equation}
\tau_0 = \dfrac{1}{2}
\sqrt{\dfrac{\dot \chi}{|\Delta|}},
\quad
\Phi(\tau) =
\dfrac{1}{2}
(\tau + \tau_0)\sqrt{s + (\tau - \tau_0)^2/4}
-
\dfrac{1}{2} s \ln s
+
s \ln \left[
\sqrt{s + (\tau - \tau_0)^2/4} + (\tau - \tau_0)/2
\right].
\end{equation}

The functions $\tilde \Psi_{\pm}(t)$ can be evaluated in the same limit using the exact solution of the Landau--Zener problem in terms of parabolic cylinder functions \cite{Z32,Iv23}. We obtain
\begin{equation}
\tilde\Psi_+(t) =
\dfrac{\sqrt{s}\Gamma(is)}{\sqrt{2\pi}}
\begin{pmatrix}
\sqrt{s} e^{i\pi/4}
\left[e^{\pi s/4}  D_{-is-1}(\tau e^{i\pi/4}) +  e^{-3\pi s/4} D_{-is-1}(\tau e^{-3i\pi/4})\right]
\\
e^{\pi s/4} D_{-is}(\tau e^{i\pi/4}) -  e^{-3\pi s/4} D_{-is}(\tau e^{-3i\pi/4})
\end{pmatrix}
\exp\left(
- i \dfrac{s}{2}\ln\dfrac{s}{e} - i \dfrac{\tau_0^2}{4}\right).
\label{Psippi}
\end{equation}
\end{widetext}
Direct analysis demonstrates that the functions $\psi_+(t)$ and $\tilde \Psi_+(t)$ defined in Eqs.~\eqref{psippi} and \eqref{Psippi} indeed have the same asymptotics in the region $\chi < \pi$ ($\tau \ll -1$). Employing these functions, one can immediately evaluate the Landau--Zener amplitudes $r(t)$ and $d(t)$ in terms of the parabolic cylinder functions,
\begin{gather}
r(t) = \psi_+^+(t)\tilde\Psi_+(t),
\quad
d(t) = \psi_+^+(t) \hat p_2 \tilde\Psi_+^*(t).
\label{rdeq}
\end{gather}
The functions $r(t)$ and $d(t)$ ~\eqref{rdeq} are substituted into Eq. (21) and the resulting dependencies $I(\chi )$ are displayed by dashed blue lines in Figs. 2-4.


\begin{thebibliography}{99}
\bibitem{FN} This behavior can easily be understood since Andreev levels "touch" the continuum of states just at the phase values $\chi =2\pi k$. Hence, at around these points quantum dynamics of Andreev levels becomes non-unitary \cite{KZ} and dissipation appears. On the other hand, sufficiently far away from these points a dissipative contribution to the current decays exponentially (cf., Eq. \eqref{diss}) and can be neglected.
\end{thebibliography}
\end{document}